\documentclass[twocolumn,showpacs,preprintnumbers,amsmath,amssymb]{revtex4}
\usepackage{graphicx, float,color}
\usepackage{bm}
\DeclareGraphicsExtensions{ps,eps,epsf}

\begin{document}

\title{Phonon anomalies in pure and underdoped \emph{R}$_{1-x}$K$_x$Fe$_2$As$_2$ (\emph{R}=Ba, Sr)\\
investigated by Raman light scattering}
\author{M.~Rahlenbeck$^1$, G. L.~Sun$^1$, D. L.~Sun$^1$, C. T.~Lin$^1$, B.~Keimer$^1$, and C.~Ulrich$^{1,2,3}$}
\affiliation{$^1$Max-Planck-Institut f\"ur Festk\"orperforschung,
Heisenbergstra\ss e 1, D-70569 Stuttgart, Germany}
\affiliation{$^2$University of New South Wales, School of Physics,
2052 Sydney, New South Wales, Australia} \affiliation{$^3$The Bragg
Institute, Australian Nuclear Science and Technology Organization,
Lucas Heights, NSW 2234, Australia}

\date{\today}

\begin{abstract}
We present a detailed temperature dependent Raman light scattering
study of optical phonons in Ba$_{1-x}$K$_{x}$Fe$_2$As$_2$
($x\sim0.28$, superconducting $T_c \sim 29$ K),
Sr$_{1-x}$K$_{x}$Fe$_2$As$_2$ ($x\sim0.15$, $T_c \sim 29$ K) and
non-superconducting BaFe$_2$As$_2$ single crystals. In all samples
we observe a strong continuous narrowing of the Raman-active Fe and
As vibrations upon cooling below the spin-density-wave transition
$T_s$. We attribute this effect to the opening of the
spin-density-wave gap. The electron-phonon linewidths inferred from
these data greatly exceed the predictions of ab-initio density
functional calculations without spin polarization, which may imply
that local magnetic moments survive well above $T_s$. A first-order
structural transition accompanying the spin-density-wave transition
induces discontinuous jumps in the phonon frequencies. These
anomalies are increasingly suppressed for higher potassium
concentrations. We also observe subtle phonon anomalies at the
superconducting transition temperature $T_c$, with a behavior
qualitatively similar to that in the cuprate superconductors.
\end{abstract}

\maketitle

The recent discovery of superconductivity in iron arsenides has
triggered a large-scale research effort to explore the physical
properties of these materials. After first reports on
LaFeAs(O$_{1-x}$F$_x$) with critical temperatures of $T_c=26$ K
\cite{Kamihara08} and $43$ K under pressure \cite{Takahashi08}, even
higher transition temperatures up to 56 K were discovered in related
compounds \cite{Wang08,Wu09}. LaFeAsO crystallizes in the
ZrCuSiAs-type crystal structure \cite{Johnson74} consisting of
alternating (LaO)$^+$ and (FeAs)$^-$ layers with one (FeAs)$^-$
layer per formula unit. In LaFeAsO, superconductivity is usually
induced by substitution of F$^-$ for O$^{2-}$, introducing electrons
into the (FeAs)$^-$ layers. It was recently shown that hole doping by substitution of La$^{3+}$
for Sr$^{2+}$ can also induce superconductivity with $T_c=25$ K
\cite{Wen08}.

More recently a second class of iron arsenide superconductors
crystallizing in the ThCr$_2$Si$_2$-type crystal structure with two
(FeAs)$^-$ layers per formula unit was \mbox{discovered
\cite{Rotter08}}. The ternary iron arsenide BaFe$_2$As$_2$ (BFA)
becomes superconducting after substitution of K$^+$ for Ba$^{2+}$,
with \mbox{$T_c=38$ K} at optimal doping \mbox{($x\sim0.4-0.5$)}
\cite{Rotter08,Chen08}. This system is a hole-doped superconductor.
Isostructural compounds with Ca \mbox{($T_c=20$ K)} \cite{Wu08}, Eu
\mbox{($T_c=32$ K) \cite{Jeevan08}}, and Sr ($T_c\sim38$ K)
\cite{Chen08a,Sasmal08} instead of Ba followed soon thereafter. A
first example of electron doping in a ternary iron arsenide, induced
by substitution of Co$^{2+}$ for Fe$^{2+}$ in BaFe$_2$As$_2$ with
\mbox{$T_c=22$ K}, was reported as well \cite{Sefat08}. In contrast
to the oxypnictides, large highly-quality single crystals are
available for this class of compounds.

Common features of the undoped iron arsenide parent compounds are a
structural phase transition from a high temperature tetragonal to a
low temperature orthorhombic or monoclinic phase, and an
antiferromagnetic spin-density-wave (SDW) transition. While for the
electron-doped compounds both phase transitions were found at
slightly different temperatures \cite{Luetkens09,Chu09}, they occur
at a same temperature $T_s$ for hole-doped BKFA \cite{Chen08}.
Undoped BaFe$_2$As$_2$ and SrFe$_2$As$_2$ show combined transitions
at $T_s\sim140$ K \cite{Rotter08b} and $T_s\sim203$ K
\cite{Tegel08}, respectively.
Doping with electrons or holes suppresses $T_s$ and induces
superconductivity at lower temperatures. In hole-doped
\emph{R}Fe$_2$As$_2$ \mbox{(\emph{R}=Ba, Sr)} strong indications
exist for a distinct doping range where the superconducting and SDW
phases coexist \cite{Chen08, Zhang08}. In the coexistence regime of
underdoped Ba$_{1-x}$K$_{x}$Fe$_2$As$_2$ (BKFA), there is evidence for a weaker structural
transition at $T_s$ without macroscopic change of the crystal
symmetry, but with an increase of microstrains, that was attributed
to a magnetically induced lattice softening in combination with
electronic phase separation \cite{Inosov09}.

Raman light scattering offers a powerful tool to detect subtle
changes in the phonon spectrum at structural or electronic phase
transitions. In this work we present a detailed temperature
dependent Raman light scattering study of the optical phonons in
superconducting BKFA and Sr$_{1-x}$K$_{x}$Fe$_2$As$_2$ (SKFA) single crystals.
We observe pronounced narrowing in the phonon linewidths at the
combined phase transition temperature $T_s$, as well as more subtle
phonon anomalies at the \mbox{superconducting phase transition
temperature $T_c$}.

\begin{figure}[H]
\includegraphics[width=8.7cm]{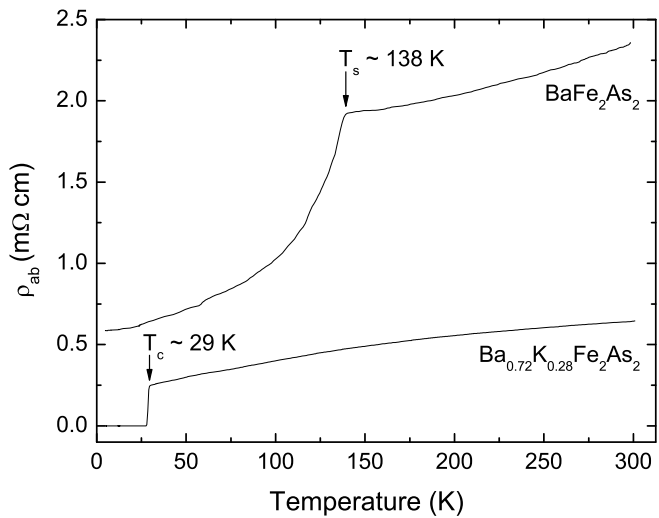}
\caption{\label{Resistivity} In-plane resistivity of BFA and BKFA single crystals determined by a standard four-probe method.}
\end{figure}

\begin{figure}[H]
\includegraphics[width=8.7cm]{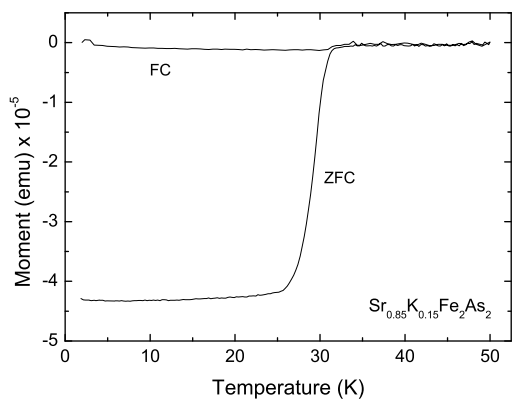}
\caption{\label{Magnetization} Field-cooled (FC) and zero-field-cooled (ZFC) magnetization of a SKFA crystal in a magnetic field of 10 Oe applied along the crystallographic $c$-axis.}
\end{figure}

The experiments were performed on thin platelets of BKFA and SKFA
single crystals $($thickness \mbox{$\sim20$ $\mu$m$)$} with the
crystallographic $c$-axis perpendicular to the surface. The samples
were grown from tin flux, as described previously
\cite{Wang09,Sun08,Park09}. They were underdoped with $x\sim0.28$
and $x\sim0.15$ (determined by energy-dispersive x-ray analysis
and inductively-coupled plasma spectroscopy), respectively.
From measurements of the electrical resistivity in the FeAs planes
(Fig. \ref{Resistivity}) and the magnetization (Fig. \ref{Magnetization})
on samples from the same batch as the ones used for the Raman measurements,
we infer superconducting transition temperatures $T_c \sim 29$ K in both cases.
In particular, the diamagnetic signal of SKFA shown in Fig. \ref{Magnetization}
is characterized by an onset of 32 K, a midpoint of 29 K, and a
10\% -- 90\% width of 4 K. Similar data were previously reported for BKFA crystals from the
same batch as ours \cite{Park09}.
The combined structural-SDW transition temperature in SKFA was
determined as $T_s \sim 178$ K by a decrease of the in-plane
resistivity on SKFA samples from the same batch \cite{Sun08}, while
$T_s \sim 75$ K was determined by neutron scattering and muon spin
rotation experiments on BKFA samples from the same batch
\cite{Park09}. Furthermore, we used a non-superconducting sample of
BFA with thickness $\sim50$ $\mu$m grown in self-flux \cite{Sun08,Ni08a}, and
$T_s \sim 138$ K inferred from the in-plane resistivity (Fig. \ref{Resistivity}).

For the Raman measurements we used a micro-Raman setup. The samples
were mounted in a helium-flow cryostat, and the spectra were taken in backscattering
geometry using the linearly polarized $632.817$ nm line of a
He$^+$/Ne$^+$-mixed gas laser for excitation. The
laser beam was focused through a $50\times$ microscope objective to
a $\sim5$ $\mu$m diameter spot on the sample surface. In order to
avoid sample heating through the laser beam, we used an incident
laser power of less than 1.5 mW. The scattered light was analyzed by
a JobinYvon LabRam single-grating spectrometer equipped with a notch
filter and a Peltier-cooled CCD camera. For each Raman spectrum an
additional calibration spectrum of a nearby neon line was measured
in order to determine the precise frequency and linewidth of the
different phonons. For data analysis, all phonon peaks were fitted to
Voigt profiles, which result from a convolution of the Lorentzian
phonon lineshape with the instrumental resolution of $\sim2$
cm$^{-1}$ (full width at half maximum).

\begin{figure}[H]
\includegraphics[width=6.7cm]{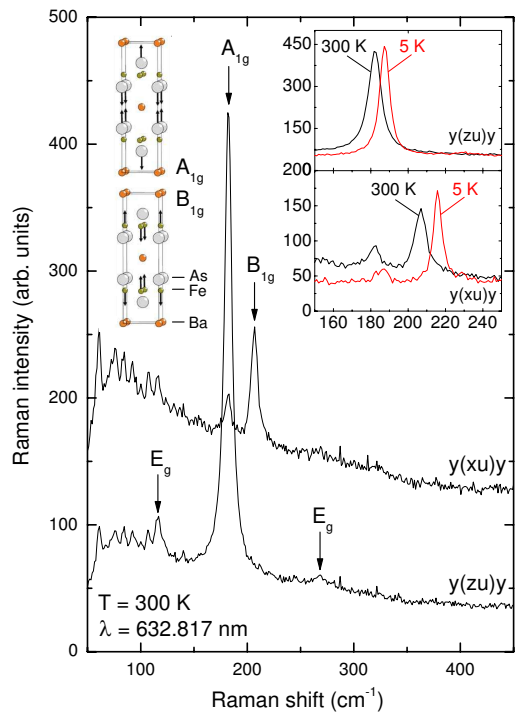}
\caption{\label{BKFA_Spektrum} Raman spectra of underdoped BKFA at
300 K in y(xu)y and y(zu)y polarization configurations.
The additional peaks below $\sim150$ cm$^{-1}$ are
artifacts of the optical setup. The insets show a comparison of the
Raman spectra at 5 and 300 K. The displacement patterns show the
Raman-active $A_{1g}$ and $B_{1g}$ modes as obtained from shell
model calculations \cite{Litvinchuk08}.}
\end{figure}

Figure \ref{BKFA_Spektrum} shows the Raman spectra of the underdoped
BKFA single crystal at $300$ K for the y(xu)y and y(zu)y
polarization configurations. Here, we use the Porto notation i(kl)j,
where i and j denote the direction of incident and scattered light
and (kl) their polarization, respectively. In order to maximize the
intensity, we did not use a polarization analyzer (u = unpolarized).
From the tetragonal ThCr$_2$Si$_2$-type crystal structure (space
group $I4/mmm$ ($D_{4h}^{17}$)) one expects four Raman-active modes
with symmetries $A_{1g}+B_{1g}+2E_{g}$. According to the mode
assignment of Litvinchuk \emph{et al.} \cite{Litvinchuk08}, the
peaks at $182$ cm$^{-1}$ and $206$ cm$^{-1}$ can be identified with
the $A_{1g}$ and $B_{1g}$ vibrations of the As- and Fe-atoms,
respectively. The $A_{1g}$ mode is strong for (zu) and weak for (xu)
polarization, respectively. The $B_{1g}$ mode cannot be observed for
(zu) polarization in accordance with the Raman selection rules. The
two $E_{g}$ modes are weak and are observed at $117$ cm$^{-1}$ and
$268$ cm$^{-1}$ in the y(zu)y polarization configuration. The insets
show the corresponding Raman spectra at 5 and 300 K. All modes show
a clear shift to lower energies upon heating from 5 to 300 K.

\begin{figure}[H]
\includegraphics[width=8.7cm]{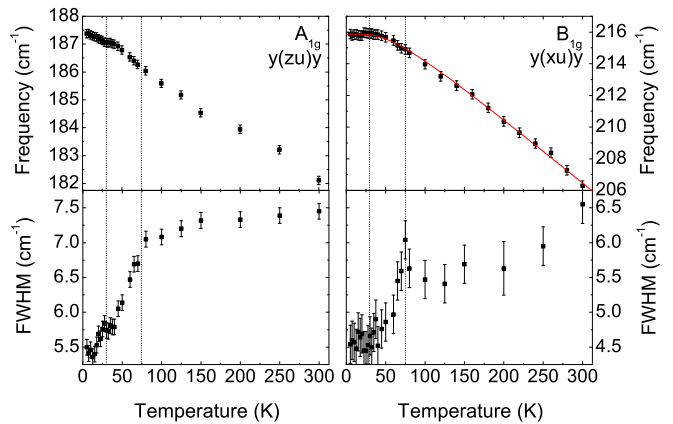}
\caption{\label{BKFA_Anomalies} Temperature dependence of the
frequency and FWHM of the $A_{1g}$ and $B_{1g}$ modes in underdoped
BKFA. The solid line is the result of a fit to the normal state data according to
anharmonic phonon decay processes (see text for details). The dashed
lines indicate the superconducting $T_c \sim 29$ K and the SDW transition
temperature $T_s\sim75$ K as obtained from Ref. \onlinecite{Park09}.}
\end{figure}

\begin{figure}[H]
\includegraphics[width=8.7cm]{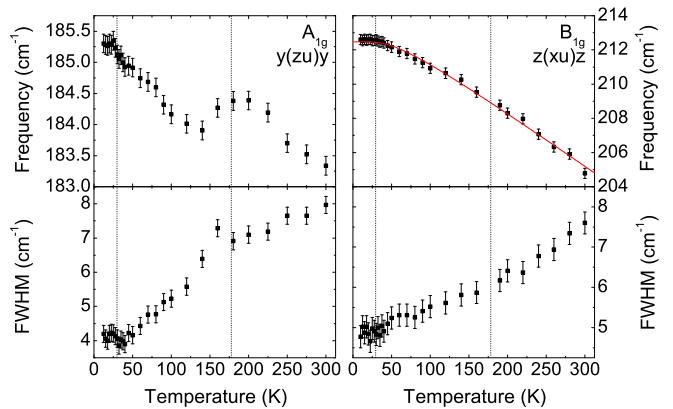}
\caption{\label{SKFA_Anomalies} Temperature dependence of the
frequency and FWHM of the $A_{1g}$ and $B_{1g}$ modes in underdoped
SKFA. The dashed lines indicate the superconducting $T_c \sim 29$ K and
the SDW transition temperature $T_s\sim 178$ K.}
\end{figure}

Figure \ref{BKFA_Anomalies} shows the frequency and full width at
half maximum (FWHM) of the $A_{1g}$ and $B_{1g}$ modes of underdoped
BKFA in the y(zu)y and y(xu)y polarization configurations as a
function of temperature. The vertical lines correspond to the
critical temperatures $T_c$ and $T_s$ for the superconducting and
the combined structural-SDW transitions, respectively. Figure
\ref{SKFA_Anomalies} shows corresponding data for underdoped SKFA
($x\sim0.15$). Compared to BKFA ($x\sim0.28$), the $A_{1g}$ and
$B_{1g}$ modes are shifted
to lower energies by $\sim1-2$ cm$^{-1}$ and $~\sim3$ cm$^{-1}$,
respectively, but the temperature evolution of the phonon parameters
is closely similar in both samples.

The solid line through the $B_{1g}$ data is the result of a fit to
an expression based on phonon-phonon interactions, i.e. the
anharmonic decay of phonons \cite{Menendez84}. For simplicity we
assumed a symmetric phonon decay according to:
\begin{eqnarray*}
\omega_{ph}(T)=-A\left(1+\frac{2a}{exp(\hbar\omega_{0}/2k_BT)-1}\right)+\omega_0,
\end{eqnarray*}

where \emph{A} is a positive constant and \emph{a}
corrects for terms arising from non-symmetric phonon decay
processes. While the frequency of the $B_{1g}$ mode follows nearly
perfectly the expression for anharmonic decay, significant
deviations from this behavior are observed for the $A_{1g}$ mode at
both $T_c$ and $T_s$. We first discuss the relatively subtle
lineshape anomalies at $T_c$, and later turn to the stronger
anomalies at $T_s$ in the context of analogous measurements on
non-superconducting BFA samples.

In contrast to the low-temperature saturation of the frequency of
the $B_{1g}$ modes, the $A_{1g}$ modes of both BKFA and SKFA exhibit
a distinct hardening by $\sim 0.3$ cm$^{-1}$ upon cooling below
$T_c$. Moreover, the linewidth of the $A_{1g}$ mode in SKFA (Fig. \ref{SKFA_Anomalies})
shows a kink in its temperature dependence at $T_c$, which can be
described as a slight superconductivity-induced broadening superposed
on a continuous decrease upon cooling below the spin-density-wave transition
at $T_s$ (to be discussed further below). While the latter trend is
already nearly saturated at $T=T_c$ in SKFA, it is much more pronounced
in BKFA, where $T_s$ is considerably lower, and overshadows the
superconductivity-induced linewidth anomaly (Fig. \ref{BKFA_Anomalies}).

According to the standard description of superconductivity-induced
self-energy anomalies of optical phonons
\cite{Zeyher90,Devereaux94}, the rearrangement of the electronic
density of states below $T_c$ is expected to induce hardening and
broadening for phonons with energies above the pair-breaking energy $2\Delta$. Experimental
observations on the cuprate high-temperature superconductors are
largely consistent with this theory. Angle-resolved photoemission
experiments on BKFA samples from the same batch as ours found two
superconducting gap energies at $\Delta_1\sim9$ meV and
$\Delta_2\sim4$ meV \cite{Evtushinsky09}, yielding pair-breaking
energies of $2\Delta_1\sim150$ cm$^{-1}$ and $2\Delta_2\sim65$
cm$^{-1}$. Assuming comparable gap values for the two BKFA and SKFA
samples, both the $A_{1g}$ and $B_{1g}$ mode energies are well above
the upper $2\Delta_1$-gap. Our observations of a hardening and
broadening in the phonon frequency and linewidth at $T_c$ are
therefore in agreement with the theoretical expectations. The
weakness of the observed self-energy renormalization indicates that
the superconductivity-induced modification of the electronic density
of states is small at the phonon energies monitored experimentally.

\begin{figure}[H]
\includegraphics[width=8.7cm]{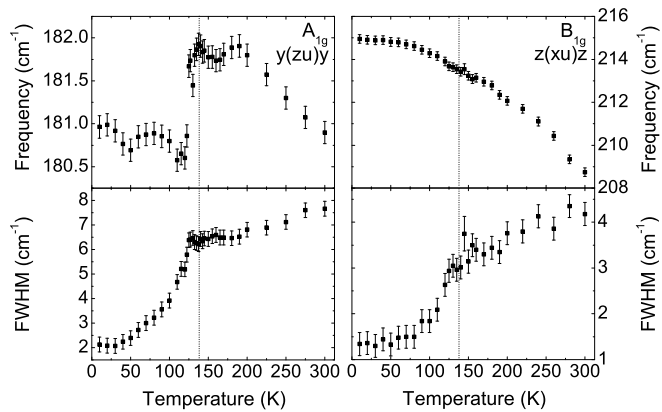}
\caption{\label{BFAk_undoped} Temperature dependence of the
frequency and FWHM of the $A_{1g}$ and $B_{1g}$ modes in undoped BFA
(grown in self-flux). The dashed lines indicate the structural-SDW
transition temperature $T_s\sim138$ K as obtained from
Fig. \ref{Resistivity}.}
\end{figure}

We now turn to the phonon anomalies at the combined structural-SDW
transition temperature $T_s$. Beginning with the BKFA sample with
the lowest $T_s$ (Fig. \ref{BKFA_Anomalies}), we note a pronounced
singularity in the temperature dependence of the
linewidths of both $A_{1g}$ and $B_{1g}$ modes at $T_s$, followed by a
continuous narrowing upon cooling. This
effect is further enhanced in the SKFA sample. The $A_{1g}$
mode, in particular, narrows by $\sim3$ cm$^{-1}$ between $T_s$ and
the lowest temperature, and softens by $\sim0.5$ cm$^{-1}$ around
$T_s$. For reference, we have carried out further experiments on
pristine BFA, the results of which are shown in Fig. \ref{BFAk_undoped}. Here
the phonon anomalies at $T_s$ are even stronger. The $A_{1g}$ mode
softens abruptly by $\sim 1$ cm$^{-1}$ at $T_s$ and narrows by more
than $\sim4$ cm$^{-1}$ between $T_s$ and the lowest temperature. The $B_{1g}$ mode frequency also shows
a small break in its temperature gradient at $T_s$, as well as a
continuous narrowing by $\sim2$ cm$^{-1}$ between $T_s$ and base
temperature. Similar behavior was reported before for CaFe$_2$As$_2$
\cite{Lemmens09}.

In assessing the physical origin of the anomalies at $T_s$, we have
to take into account the modification of the lattice structure and
the opening of the SDW gap at this transition. As shown by several
authors \cite{Huang08,Ni08b,Krellner08}, the structural transition in
the ternary iron arsenides tends to be of first order. The
discontinuous jump in the frequency of the $A_{1g}$ mode at $T_s$ is
consistent with this behavior and can be attributed to the
contraction of the lattice at the structural transition
\cite{Inosov09}. The reduction of this discontinuity in the BKFA
sample with the lowest $T_s$ is in agreement with earlier
experiments on BKFA samples from the same batch \cite{Inosov09},
where the structural phase transition was found to be
macroscopically suppressed. Note that recent Raman
scattering measurements on electron-doped
Ba(Fe$_{1-x}$Co$_x$)$_2$As$_2$ single crystals \cite{Chauviere09}
revealed a splitting of the degenerate $E_g$ in-plane modes
at the tetragonal-to-orthorhombic transition.

On the other hand, the pronounced and continuous narrowing of both
$A_{1g}$ and $B_{1g}$ modes below $T_s$ cannot be understood as a
consequence of the modification of the lattice structure and
associated phonon-phonon interactions alone. It can, however, be
understood if the difference between the linewidths above and below
$T_s$ is attributed to the electron-phonon interaction, which
becomes inoperative at low temperatures due to the opening of the
SDW gap \cite{Hu08}. This scenario is further
supported by infrared experiments on EuFe$_2$As$_2$ single crystals
\cite{Wu09a}, where the interaction of the Fe-As lattice vibration
$E_u$ with the electronic background was found to be reduced below
the SDW transition. The continuous, second-order-like behavior of
the linewidth can be reconciled with the abrupt jump in the phonon
frequency either if separate first-order structural and second-order
SDW transitions take place at slightly different temperatures that
are not resolved in the experiment, or if the single, combined
transition is weakly first-order so that the jump in linewidth is
below the detection limit.

In any case, the strong reduction of the linewidths of both $A_{1g}$
and $B_{1g}$ modes below $T_s$  points to a substantial influence of
the electron-phonon interaction on the lattice dynamics in the iron
arsenides. Ab-initio density functional calculations predict that
the $A_{1g}$ mode, which corresponds to a vibration of the arsenic
ions along the c-axis, shows the strongest electron-phonon coupling
\cite{Boeri08,Yildirim09}, in qualitative agreement with
our observations. However, calculations without spin polarization
fail to account quantitatively for the experimentally observed
linewidths. The predicted \cite{Boeri} density of states at the
Fermi level, $N(0) \sim 4.42$/eV, and total \cite{Boeri08}
electron-phonon coupling parameter (summed over the phonon branches
and averaged on the Brillouin zone), $\lambda \sim 0.21$, yields a
rough estimation for the FWHM \cite{Boeri08,Keimer} of
\mbox{$\frac{1}{2} \pi N(0) \frac{\lambda}{15} \omega^2 \sim 0.4$
cm$^{-1}$}, about an order of magnitude smaller than observed for
the $A_{1g}$ mode above the SDW transition. This discrepancy is in
line with earlier reports of differences between the calculated and
observed phonon density of states around the
$A_{1g}$ mode energy \cite{fukuda08,reznik09,Mittal08,Mittal09}. Since the ab-initio calculations also show that the
electron-phonon coupling greatly increases in the SDW state
\cite{Yildirim09,Boeri09}, this may be a manifestation of local Fe
moments that survive well above the SDW transition.

In conclusion, we have shown that the electron-phonon coupling
results in significant phonon anomalies at both the superconducting
and spin-density-wave transitions of BKFA and SKFA. The anomalies at
$T_c$ are small, presumably because the phonon energies exceed the
pair breaking energies, and consistent with the conventional theory
of superconductivity-induced phonon self-energy modifications. The
anomalies at $T_s$ are much larger and qualitatively consistent with
the opening of the spin-density-wave gap. A quantitative description
of these anomalies requires theoretical work beyond the density
functional calculations thus far reported.

We would like to thank L. Boeri and M. Le Tacon for fruitful
discussions and A. Schulz for technical support during the Raman
experiments.


\begin{thebibliography}{}

\bibitem{Kamihara08} Y. Kamihara, T. Watanabe, M. Hirano, and H. Hosono, J. Am. Chem. Soc. \textbf{130}, 3296 (2008).

\bibitem{Takahashi08} H. Takahashi, K. Igawa, K. Arii, Y. Kamihara, M. Hirano, and H. Hosono, Nature \textbf{453}, 376 (2008).

\bibitem{Wang08} C. Wang, L. Li, S. Chi, Z. Zhu, Z. Ren, Y. Li, Y. Wang, X. Lin,
Y. Luo, S. Jiang, X. Xu, G. Cao, and Z. Xu, Europhys. Lett.
\textbf{83}, 67006 (2008).

\bibitem{Wu09} G. Wu, Y. L. Xie, H. Chen, M. Zhong, R. H. Liu, B. C. Shi, Q. J. Li,
X. F. Wang, T. Wu, Y. J. Yan, J. J. Ying, and X.H. Chen, J. Phys.:
Condens. Matter \textbf{21}, 142203 (2009).

\bibitem{Johnson74} V. Johnson and W. Jeitschko, J. Solid State Chem. \textbf{11}, 161 (1974).

\bibitem{Wen08} H.-H. Wen, G. Mu, L. Fang, H. Yang, and X. Zhu, Europhys. Lett. \textbf{82}, 17009 (2008).

\bibitem{Rotter08} M. Rotter, M. Tegel, and D. Johrendt, Phys. Rev. Lett. \textbf{101}, 107006 (2008).

\bibitem{Chen08} H. Chen, Y. Ren, Y. Qiu, W. Bao, R. H. Liu, G. Wu, T. Wu, Y. L. Xie, X. F. Wang,
Q. Huang, and X. H. Chen, Europhys. Lett. \textbf{85}, 17006 (2009).

\bibitem{Wu08} G. Wu, H. Chen, T. Wu, Y. L. Xie, Y. J. Yan, R. H. Liu, X. F. Wang,
J. J. Ying, and X. H. Chen, J. Phys.: Condens. Matter \textbf{20},
422201 (2008).

\bibitem{Jeevan08} H. S. Jeevan, Z. Hossain, D. Kasinathan, H. Rosner, C. Geibel, and P. Gegenwart, Phys. Rev. B \textbf{78}, 092406 (2008).

\bibitem{Chen08a} G. F. Chen, Z. Li, G. Li, W.-Z. Hu,
J. Dong, J. Zhou, X.-D. Zhang, P. Zheng, N.-L. Wang, and J.-L. Luo,
Chin. Phys. Lett. \textbf{25}, 3403 (2008).

\bibitem{Sasmal08} K. Sasmal, B. Lv, B. Lorenz, A. M. Guloy, F. Chen, Y.-Y. Xue, and C.-W. Chu, Phys. Rev. Lett. \textbf{101}, 107007 (2008).

\bibitem{Sefat08} A. S. Sefat, R. Jin, M. A. McGuire, B. C. Sales, D. J. Singh, and D. Mandrus, Phys. Rev. Lett. \textbf{101}, 117004 (2008).

\bibitem{Luetkens09} H. Luetkens, H.-H. Klauss, M. Kraken, F. J. Litterst, T. Dellmann, R. Klingeler, C. Hess,
R. Khasanov, A. Amato, C. Baines, M. Kosmala, O. J. Schumann, M.
Braden, J. Hamann-Borrero, N. Leps, A. Kondrat, G. Behr, J. Werner,
and B. B\"uchner, Nature Materials \textbf{8}, 305 (2009).

\bibitem{Chu09} J.-H. Chu, J. G. Analytis, Ch. Kucharczyk, and I. R. Fisher, Phys. Rev. B \textbf{79}, 014506 (2009).

\bibitem{Rotter08b} M. Rotter, M. Tegel, D. Johrendt, I. Schellenberg, W. Hermes, and R. P\"ottgen, Phys. Rev. B \textbf{78}, 020503(R) (2008).

\bibitem{Tegel08} M. Tegel, M. Rotter, V. Wei\ss,
F. M. Schappacher, R. P\"ottgen, and D. Johrendt, J. Phys.: Condens.
Matter \textbf{20}, 452201 (2008).

\bibitem{Zhang08} Y. Zhang, J. Wei, H. W. Ou, J. F. Zhao, B. Zhou, F. Chen, M. Xu, C. He, G. Wu, H.
Chen, M. Arita, K. Shimada, H. Namatame, M. Taniguchi, X. H. Chen,
and D. L. Feng, Phys. Rev. Lett. {\bf 102}, 127003 (2009).

\bibitem{Inosov09} D. S. Inosov, A. Leineweber, X. Yang, J. T. Park, N. B. Christensen, R. Dinnebier, G. L. Sun,
Ch. Niedermayer, D. Haug, P. W. Stephens, J. Stahn, O. Khvostikova,
C. T. Lin, O. K. Andersen, B. Keimer, and V. Hinkov,
Phys. Rev. B \textbf{79}, 224503 (2009).

\bibitem{Wang09} X.F. Wang, T. Wu, G. Wu, H. Chen, Y.L. Xie, J.J. Ying, Y.J. Yan, R.H. Liu, and X.H. Chen, Phys. Rev. Lett. {\bf 102}, 117005 (2009).

\bibitem{Sun08} G. L. Sun, D. L. Sun, M. Konuma, P. Popovich, A. Boris, J. B. Peng,
K.-Y. Choi, P. Lemmens, and C. T. Lin, arXiv: 0901.2728v2 (2009).

\bibitem{Park09} J. T. Park, D. S. Inosov, Ch. Niedermayer, G. L. Sun, D. Haug, N. B. Christensen, R. Dinnebier, A. V. Boris,
A. J. Drew, L. Schulz, T. Shapoval, U. Wolff, V. Neu, X. Yang, C. T.
Lin, B. Keimer, and V. Hinkov, Phys. Rev. Lett. \textbf{102}, 117006
(2009).

\bibitem{Ni08a} N. Ni, S.L. Bud'ko, A. Kreyssig, S. Nandi, G.E. Rustan, A.I. Goldman, S. Gupta, J.D. Corbett, A.
Kracher and P.C. Canfield, Phys. Rev. B {\bf 78}, 014507 (2008).

\bibitem{Litvinchuk08} A. P. Litvinchuk, V. G. Hadjiev, M. N. Iliev, B. Lv, A. M. Guloy, and C. W. Chu, Phys. Rev. B \textbf{78}, 060503(R) (2008).

\bibitem{Menendez84} J. Men\'{e}ndez and M. Cardona, Phys. Rev. B \textbf{29}, 2051 (1984).

\bibitem{Zeyher90} R. Zeyher and G. Zwicknagl, Z. Phys. B - Condens. Matter \textbf{78}, 175 (1990).

\bibitem{Devereaux94} T. P. Devereaux, Phys. Rev. B \textbf{50}, 10287 (1994).

\bibitem{Evtushinsky09} D. V. Evtushinsky, D. S. Inosov, V. B. Zabolotnyy, A. Koitzsch, M. Knupfer, B. B\"uchner, M. S. Viazovska,
G. L. Sun, V. Hinkov, A. V. Boris, C. T. Lin, B. Keimer, A.
Varykhalov, A. A. Kordyuk, and S. V. Borisenko, Phys. Rev. B
\textbf{79}, 054517 (2009).

\bibitem{Lemmens09} K.-Y. Choi, D. Wulferding, P. Lemmens, N. Ni, S. L. Bud'ko, and P. C. Canfield, Phys. Rev. B \textbf{78}, 212503 (2008).

\bibitem{Huang08} Q. Huang, Y. Qiu, W. Bao, M. A. Green, J. W. Lynn, Y. C. Gasparovic, T. Wu, G.
Wu, and X. H. Chen, Phys. Rev. Lett. {\bf 101}, 257003 (2008).

\bibitem{Ni08b} N. Ni, S. Nandi, A. Kreyssig, A. I. Goldman, E. D. Mun, S. L. Bud'ko, and P. C. Canfield, Phys. Rev. B \textbf{78}, 014523 (2008).

\bibitem{Krellner08} C. Krellner, N. Caroca-Canales, A. Jesche, H. Rosner, A. Ormeci, and C. Geibel, Phys. Rev. B \textbf{78}, 100504(R) (2008).

\bibitem{Chauviere09} L. Chauvi\`{e}re, Y. Gallais,
M. Cazayous, A. Sacuto, M. A. M\'{e}asson, D. Colson and A. Forget,
arXiv: 0906.4569v1 (2009).

\bibitem{Hu08} W. Z. Hu, J. Dong, G. Li, Z. Li, P. Zheng, G. F. Chen, J. L. Luo, and N. L. Wang, Phys. Rev. Lett. \textbf{101}, 257005 (2008).

\bibitem{Wu09a} D. Wu, N. Bari\v{s}i\'{c}, N.
Drichko, S. Kaiser, A. Faridian, M. Dressel, S. Jiang, Z. Ren, L. J.
Li, G. H. Cao, Z. A. Xu, H. S. Jeevan, and P. Gegenwart, Phys. Rev.
B {\bf 79}, 155103 (2009).

\bibitem{Boeri08} L. Boeri, O. V. Dolgov, and A. A. Golubov, Phys. Rev. Lett. \textbf{101}, 026403 (2008).

\bibitem{Yildirim09} T. Yildirim, Phys. Rev. Lett.
\textbf{102}, 037003 (2009); Physica C \textbf{469}, 425 (2009).

\bibitem{Boeri} L. Boeri, private communication.

\bibitem{Keimer} Note that the factor of 15 reflects the number of
phonon branches in BaFe$_2$As$_2$.

\bibitem{fukuda08} T. Fukuda, A. Q. R. Baron, Sh.-I. Shamoto, M. Ishikado,
H. Nakamura, M. Machida, H. Uchiyama, S. Tsutsui, A. Iyo, H. Kito,
J. Mizuki, M. Arai, H. Eisaki, and H. Hosono, J. Phys. Soc. Jpn.
{\bf 77}, 103715 (2008).

\bibitem{reznik09} D. Reznik, K. Lokshin, D. C. Mitchell, D. Parshall, W. Dmowski, D. Lamago, R. Heid,
K.-P. Bohnen, A. S. Sefat, M. A. McGuire, B. C. Sales, D. G.
Mandrus, A. Asubedi, D. J. Singh, A. Alatas, M. H. Upton, A. H.
Said, Yu. Shvyd'ko, and T. Egami, arXiv: 0810.4941 (2008).

\bibitem{Mittal08} R. Mittal, Y. Su, S. Rols, T. Chatterji, S. L. Chaplot, H. Schober, M. Rotter, D. Johrendt, and Th.
Brueckel, Phys. Rev. B \textbf{78}, 104514 (2008).

\bibitem{Mittal09} M. Zbiri, H. Schober, M. R. Johnson, S. Rols, R. Mittal, Y. Su, M. Rotter, and D.
Johrendt, Phys. Rev. B \textbf{79}, 064511 (2009).

\bibitem{Boeri09} L. Boeri, O. V. Dolgov, A. A. Golubov, Physica C {\bf 469}, 628 (2009).


\end{thebibliography}
\end{document}